\documentclass[aps,showpacs,nofootinbib,superscriptaddress]{revtex4-1}
\usepackage{epsf}
\usepackage{graphicx}
\usepackage{amsmath}
\usepackage{url}  
\def\n#1e#2n{#1\times10^{#2}}
\def\g#1{\gamma_{#1}}
\makeatletter
\def\slashchar#1{{\mathpalette\c@ncel{#1}}} 
\makeatother
\def\vsl{\slashchar{v}}

\newcommand{\bea}{\begin{eqnarray}}
\newcommand{\eea}{\end{eqnarray}}
\newcommand{\be}{\begin{equation}}
\newcommand{\ee}{\end{equation}}

\begin{document}


\title{Triply Heavy Baryons and Heavy Quark Spin Symmetry}

\author{J. M.~Flynn}\affiliation{School of Physics and
  Astronomy, University of Southampton, Highfield, Southampton
  SO17~1BJ, UK}
\author{E.~Hern\'andez}\affiliation{Departamento de F\'\i sica Fundamental e
IUFFyM, Universidad de Salamanca, E-37008 Salamanca, Spain.}
\author{J.~Nieves}\affiliation{Instituto de F\'\i sica Corpuscular
(IFIC), Centro Mixto CSIC-Universidad de Valencia, Institutos de
Investigaci\'on de Paterna, Apartado 22085, E-46071 Valencia, Spain.}

\date{\today}

\begin{abstract}
We study the semileptonic $b\to c$ decays of the lowest-lying triply-heavy
baryons made from $b$ and $c$ quarks in the limit $m_b, m_c \gg
\Lambda_\mathrm{QCD}$ and close to the zero recoil point. The separate
heavy quark spin symmetries strongly constrain the matrix elements,
leading to single form factors for $ccb\to ccc$, $bbc\to ccb$, and 
 $bbb\to bbc$ baryon decays. We also study the effects on these systems
of using a $Y$-shaped confinement potential, as suggested by lattice
QCD  results for the interaction between three static quarks.
\end{abstract}

\maketitle

\section{Introduction}

Triply heavy baryons are systems of great theoretical interest, since
they may serve to better understand the interaction among heavy quarks
in an environment free of valence light quarks. Besides, being
baryonic analogues of heavy quarkonium, they might yield sharp tests
for QCD. Studying these baryons will be also very useful for
understanding the three quark static potential.

With no experimental information available on these systems, previous
studies have concentrated on their spectrum. To our knowledge the
first such study was carried out in 1980~\cite{Hasenfratz:1980ka}
using a QCD-motivated bag model (BM). A mass formula was derived by
Bjorken in Ref.~\cite{Bjorken} providing predictions for the masses
that were larger than those found in Ref.~\cite{Hasenfratz:1980ka}. In
Ref.~\cite{Bjorken} the possibility for discovery of the
$\Omega_{ccc}$ state was also discussed. More recently there has been
other phenomenological mass determinations that include
nonrelativistic constituent quark model (NRCQM)
calculations~\cite{silvestre96,Vijande:2004at,Roberts:2007ni}, the
relativistic three quark model (RTQM) evaluation of
Ref.~\cite{Martynenko:2007je}, or the Regge approach in
Ref.~\cite{Guo:2008he}. More fundamental approaches to the subject
include the potential nonrelativistic QCD (pNRQCD) studies of
Refs.~\cite{Brambilla:2005yk,Brambilla:2009cd} or the QCD sum rule
(QCDSR) evaluation of Ref.~\cite{Zhang:2009re}.  In
Ref.~\cite{Jia:2006gw} the leading order (LO) pNRQCD result of
Ref.~\cite{Brambilla:2005yk} is used\footnote{This coincides with the
$1/r$, or Coulomb, interaction that comes from one-gluon exchange},
while a mass calculation that includes next-to-next-to-leading order
within the same framework has just appeared~\cite{felipe11}. The mass
of the triply-heavy baryon $\Omega_{bbb}$ has been also recently calculated
in Lattice QCD (LQCD) using $2+1$ flavours of light sea
quarks~\cite{Meinel:2010pw}.

Triply-charmed baryon production in the $e^+e^-$ reaction was analyzed
in Ref.~\cite{baranov} with the result that the predicted production
rate was very small. Better perspectives for production are expected
at the LHC due to its high luminosity. First estimates of the cross
section production at LHC were evaluated in
Refs.~\cite{saleev,gomshi03,gomshi05,gomshi06}. A recent
evaluation~\cite{Chen:2011mb} finds that around $10^4$--$10^5$ events of
triply heavy baryons, with $ccc$ and $ccb$ quark content, can be
accumulated for $10$\,fb of integrated luminosity. The authors of this
latter work conclude that it is quite likely triply heavy baryons
would be discovered at LHC. With this in mind, study of their
properties beyond spectroscopy seems timely.

In this work, we will study the lowest lying ($J^\pi=1/2^+,3/2^+$)
triply heavy baryons composed of $b$ and $c$ quarks\footnote{In what
  follows we will denote by $\Xi$ the baryons with spin 1/2, while we
  will use $\Xi^*$ and $\Omega^*$ for the spin 3/2 ones. }. Heavy
quark spin symmetry (HQSS) is of particular interest to study these
systems. HQSS is an approximate symmetry of QCD in the limit $m_b,m_c
\gg \Lambda_{\rm QCD}$, and has proved to be an extremely useful tool
when dealing with heavy hadrons~\cite{white91, hep-ph/9204238,MWbook}.
This symmetry amounts to the decoupling of the heavy quark
spins~\cite{white91,hep-ph/9204238}. In
Ref.~\cite{SanchisLozano:1994vh} it is argued that this symmetry
cannot be considered as asymptotically valid in heavy-heavy states,
since the momentum exchange between two heavy quarks might be much
larger than $\Lambda_{\rm QCD}$, in sharp contrast to the situation
for heavy-light systems. For mesons with two equal-mass heavy quarks,
Ref.~\cite{SanchisLozano:1994vh} argues that the hyperfine splitting
$\Delta m$ scales as $\Delta m\approx m_Q\,\alpha^4_s(m_Q)$ for
sufficiently large heavy-quark mass $m_Q$ and thus asymptotically
increases with $m_Q$. However, $\frac{\Delta m}{m}\approx
\alpha^4_s(m_Q)$ still approaches zero as $m_Q$ tends to infinity and
the hyperfine splitting becomes negligible compared to the total mass.
Moreover, the linear behavior is estimated in
Ref.~\cite{SanchisLozano:1994vh} to take over from the
$\alpha^4_s(m_Q)$ logarithmic fall off for heavy-quark masses in the
region of $10\,\mathrm{GeV}$. Hence for systems with two heavy quarks,
with masses below $10\,\mathrm{GeV}$, HQSS should still be valid.
That being the case, HQSS should be a useful approximate symmetry to
address baryons made out of three $c$ and/or $b$ heavy quarks, as we
aim to do in this work.

The study of baryons requires the solution of the three-body problem.
In the past we have made extensive use of HQSS constraints and have
developed a simple variational scheme to find masses and wave
functions of single~\cite{Albertus:2003sx} and
double~\cite{Albertus:2006wb} heavy baryons. We have used the
resulting wave functions to study their semileptonic $b\to
c$~\cite{Albertus:2006wb,Albertus:2004wj,Hernandez:2007qv,Albertus:2009ww}
and $c\to s,d$~\cite{Albertus:2011xz} decays. The separate heavy quark
spin symmetries strongly constrain the matrix elements and, in the
limit $m_b,m_c \gg \Lambda_{\rm QCD}$ and close to the zero-recoil
point, they lead to single form factors for all these decays.

Here, we extend our scheme to study triply heavy baryons. We derive
for the first time HQSS relations for their semileptonic $b\to c$
decays  from which we can make approximate, but model
independent, predictions for some decay width ratios. We give absolute
values of the semileptonic $b\to c$ decay widths, as well.  We also study
the effects in these baryons of considering a LQCD inspired three-body
confinement potential (denoted as $Y$ in ~\cite{Hasenfratz:1980ka})
instead of the commonly used one, obtained from the sum of two-body 
quark-quark terms. 

\section{Spin Symmetry}

We will consider decays induced by the semileptonic weak decay of a
$b$ quark to a $c$ quark. Near the zero recoil point, the velocities
of the initial and final baryons are approximately the same. If the
momenta of the initial and final baryons are $p_\mu = m v_\mu$ and
$p'_\mu = m'v'_\mu= m'v_\mu + k_\mu$ respectively, then $k$ will be a
small residual momentum near the zero-recoil point. For the initial
baryon at rest we have that $k\cdot v = E'-m'$. For a small final
momentum this is approximately given by $\vec p\,'^2/2m'$ and then is
$\mathcal{O}(1/m')$ close to zero recoil. We will work near
zero-recoil and thus neglect $v\cdot k$ below.

The consequences of spin symmetry for weak matrix elements can be
derived using the ``trace formalism''~\cite{MWbook,Falk}. The scheme
advocated here is similar to that employed in Ref.~\cite{Flynn:2007qt}
to study the semileptonic $bc$ to $cc$ baryon decays. To represent
baryons with three heavy quarks we will use wave functions comprising
tensor products of Dirac matrices and spinors. For $Q_1Q_1Q_2$ baryons
containing two heavy quarks $Q_1$ and a distinct heavy quark $Q_2$, we
have:
\begin{align}
\label{eq:Xi}
 \Xi_{Q_1Q_1Q_2} &=
 \left[\frac{(1+\vsl)}2 \g\mu\frac{(1-\vsl)}2\right]_{\alpha\beta}
 \left[\frac1{\sqrt3}
 (v^\mu+\gamma^\mu)\g5  u(v,r)\right]_\gamma\\
\label{eq:Xistar}
\Xi^*_{Q_1Q_1Q_2} &=
 \left[\frac{(1+\vsl)}2 \g\mu\frac{(1-\vsl)}2\right]_{\alpha\beta}
 u^\mu_\gamma(v,r)
\end{align}
where we have indicated Dirac quark indices $\alpha$, $\beta$ and $\gamma$
explicitly on the right-hand sides and $r$ is a helicity label for the
baryon\footnote{Note that the two identical heavy quarks $Q_1$ 
can only be in a symmetric spin 1 state. The structure 
\begin{equation}
\left[\frac{(1+\vsl)}2 \g\mu\frac{(1-\vsl)}2\right]
\end{equation}
guarantees that the spin of the first two heavy quarks is coupled to $1$
(see for instance Refs.~ \cite{hep-ph/9204238,MWbook}). On the other
hand, the spin-$1/2$ spinor $\left[\frac1{\sqrt3}
 (v^\mu+\gamma^\mu)\g5  u(v,r)\right]$ is discussed in
Ref.~\cite{SLAC-PUB-5689}.}. For the
$\Xi^*$ states, $u^\mu_\gamma(v,r)$ is a Rarita-Schwinger spinor. For
the baryon containing three heavy quarks of the same flavour, we use:
\begin{equation}
\label{eq:Omegastar}
\Omega^*_{QQQ} = \frac1{\sqrt3}
 \left[\frac{(1+\vsl)}2 \g\mu\frac{(1-\vsl)}2\right]_{\alpha\beta}
 u^\mu_\gamma(v,r)
\end{equation}
These wave functions can be considered as matrix elements of the form
$\langle0 | {Q_1}_\alpha \bar{Q_1^c}_\beta {Q_2}_\gamma |
B_{Q_1Q_1Q_2}\rangle$ where $\bar{Q^c}=Q^T C$ with $C$ the
charge-conjugation matrix. In each case we couple two quarks of the
same flavour in a symmetric spin-$1$ state in the first factor and
combine with a spinor for the third quark. Under a Lorentz
transformation, $\Lambda$, and heavy quark spin transformations
$S_Q$, a wave function of the form $\Gamma_{\alpha\beta}\,
{\cal U}_\gamma$, with ${\cal U}= \frac{1}{\sqrt3} (v^{\mu} +
\gamma^\mu) \g5 u$ or $u^\mu $, transforms as:
\begin{equation}
\label{eq:spintransfs}
\Gamma\,{\cal U} \to S(\Lambda) \Gamma S^{-1}(\Lambda)\; S(\Lambda)
	{\cal U},
\quad
\Gamma\,{\cal U} \to S_{Q_1} \Gamma S_{Q_1}^\dagger \, S_{Q_2} {\cal U}.
\end{equation}
The $Q_1Q_1Q_2$ states have normalization $\bar {\cal U} {\cal U}
\mathrm{Tr}(\Gamma \overline\Gamma)$, while for $QQQ$ the
normalization is $\bar {\cal U} {\cal U}\, \mathrm{Tr}(\Gamma
\overline\Gamma) + 2\, \bar {\cal U}\, \Gamma \overline\Gamma \,{\cal
  U}$ (which can be understood by counting quark contractions). We
define $\overline\Gamma = \gamma^0 \Gamma^\dagger \gamma^0$ as usual
and our spinors satisfy $\bar u u = 2 m$, $\bar u^\mu u_\mu = - 2 m$
where $m$ is the mass of the state.

We construct amplitudes for semileptonic decays determined by matrix
elements of the weak current $j^\mu = \bar c \gamma^\mu(1-\g5) b$. The
operator $\bar c J^\mu b$, where $J^\mu=\gamma^\mu(1-\g5)$, would be
invariant under heavy-quark spin transformations if $J^\mu$
transformed as $J^\mu \to S_c J^\mu S_b^\dagger$. Thus, we can build
matrix elements respecting the heavy quark spin symmetry by
constructing quantities which would be invariant under the same
assumption. We observe that $j^\mu$ can be rewritten as $j^\mu =
-\bar{b^c}\gamma^\mu(1+\g5)c^c$ and note that $\bar{b^c}J^{c\,\mu}
c^c$, where $J^{c\,\mu}=-\gamma^\mu(1+\g5)$, would be invariant if
$J^{c\,\mu}\to S_b J^{c\,\mu} S_c^\dagger$.

For the transitions $\Xi^{(*)}_{ccb} \to \Omega^*_{ccc}$, the matrix
element respecting heavy quark symmetry is, up to a scalar function
of the product of velocities, $w=v\cdot v'$,
\begin{equation}
\label{eq:ME1}
\langle \Omega^*_{ccc},v,k,r'|j^\mu(0)|\Xi^{(*)}_{ccb},v,r\rangle 
 =\bar {\cal U}'(v,k,r') J^\mu {\cal U}(v,r)
   \mathrm{Tr}[\Gamma_{ccb} \overline\Gamma_{ccc}]
  + \bar {\cal U}'(v,k,r') \Gamma_{ccb}\overline\Gamma_{ccc} J^\mu {\cal
 U}(v,r)
\end{equation}
where $r$ and $r'$ are the helicities of the initial and final states,
and we use the standard relativistic normalization for hadronic states.
Terms with a factor of $\vsl$ can be omitted because of the equations
of motion ($\vsl u=u$, $\vsl\Gamma=\Gamma$, $\g\mu u^\mu=0$, $v_\mu
u^\mu=0$), while terms with $\slashchar k$ will always lead to
contributions proportional to $v\cdot k$ which is set to $0$ at the
order we are working. We also make use of the exact relation $\bar u'
\slashchar k u =0$ and the approximate ones $\bar u' \g j u = \bar u'
v_j u$, $\bar u' \g5 u = 0$, and $\bar u' \slashchar k \g\mu \g5 u = -
\bar u' \slashchar k v_\mu \g5 u$ valid close to zero recoil.

For the transitions $\Xi^{(*)}_{bbc} \to \Xi^{(*)}_{ccb}$ the matrix
element is
\begin{equation}
\label{eq:ME2}
\langle\Xi^{(*)}_{ccb},v,k,r'|j^\mu(0)|\Xi^{(*)}_{bbc},v,r\rangle 
 = \bar {\cal U}'(v,k,r') \Gamma_{bbc}J^{c\,\mu}\overline\Gamma_{ccb}
 {\cal U}(v,r),
\end{equation}
while for the transitions $\Omega^*_{bbb} \to \Xi^{(*)}_{bbc}$, the matrix
element respecting heavy quark symmetry now reads
\begin{equation}
\label{eq:ME3}
\langle\Xi^{(*)}_{bbc},v,k,r'|j^\mu(0)|\Omega^*_{bbb},v,r\rangle 
 =\bar {\cal U}'(v,k,r') J^\mu {\cal U}(v,r)
   \mathrm{Tr}[\Gamma_{bbb} \overline\Gamma_{bbc}]
  + \bar {\cal U}'(v,k,r') \Gamma_{bbb}\overline\Gamma_{bbc} J^\mu {\cal
 U}(v,r)
\end{equation}

Close to zero recoil, and within the approximations mentioned above, our results for the transition matrix elements, apart from irrelevant
global phases, are:
\begin{eqnarray}
\label{eq:transitions}
\Xi_{ccb}\to \Omega^*_{ccc}  &&\hspace{1cm}
 2\eta\, \bar u'^\mu u
 \label{eq:tran1}\\
\Xi^*_{ccb}\to \Omega^*_{ccc}  &&\hspace{1cm}
-{\sqrt3}\eta\, \bar u'^\lambda \gamma^\mu(1-\g5) u_{\lambda}  \label{eq:tran2}\\
 \nonumber\\
\Xi_{bbc}\to \Xi_{ccb}  &&\hspace{1cm}
 -\chi \bar u'\big(\gamma^\mu-\frac53\gamma^\mu\g5\big) u  \label{eq:tran3}\\
\Xi_{bbc}\to \Xi^*_{ccb}  &&\hspace{1cm}
 -\frac2{\sqrt3} \chi \bar u'^\mu u  \label{eq:tran4}\\
\Xi^*_{bbc}\to \Xi_{ccb}  &&\hspace{1cm}
 -\frac2{\sqrt3} \chi \bar u' u^\mu \label{eq:tran5} \\
\Xi^*_{bbc}\to \Xi^*_{ccb}  &&\hspace{1cm}
 -2 \chi \bar u'^\lambda \gamma^\mu(1-\g5) u_{\lambda}
 \label{eq:tran6}\\\nonumber\\
\Omega^*_{bbb}\to \Xi_{bbc}  &&\hspace{1cm}
 2\xi\, \bar u' u^\mu \label{eq:tran7}\\
\Omega^*_{bbb}\to \Xi^*_{bbc}  &&\hspace{1cm}
-{\sqrt3}\xi\, \bar u'^\lambda \gamma^\mu(1-\g5) u_{\lambda} \label{eq:tran8}
\end{eqnarray}
where the factors $\eta(w)$, $\chi(w)$ and $\xi(w)$ are the Isgur-Wise
functions that depend on $w=v\cdot v'$ and that we expect to be close
to 1 at zero recoil ($w=1$). In fact in the limit $m_c=m_b$ they would
be exactly $1$ at zero recoil. To check that assertion let us consider
an $SU(2)$ symmetry under which the $c$ and $b$ quarks transform as a
doublet and the four states $\Omega^*_{ccc}$, $\Xi^*_{ccb}$,
$\Xi^*_{bbc}$, and $\Omega^*_{bbb}$ form a quadruplet. We will
consider all heavy quark spins aligned, that is to say we will place
the four baryons in the state with maximum third component of spin,
$J_z=+3/2$. The $\mu=0$ component of the vector part of the transition
operator $j^\mu$ would then be $I_+=c^\dagger b$, which is the raising
operator in the Fock space for this flavour $SU(2)$ symmetry. Assuming this
symmetry, and taking into account the state normalization, we will
have at zero recoil
\begin{eqnarray}
\label{eq:norm1}
\sqrt3=\frac1{2m}\langle \Omega^*_{ccc}|c^\dagger
b|\Xi^*_{ccb}\rangle&=&\sqrt3\,\eta(1) \\
\label{eq:norm2}
2=\frac1{2m}\langle \Xi^*_{ccb}|c^\dagger b|\Xi^*_{bbc}\rangle&=&2\,\chi(1) \\
\label{eq:norm3}
\sqrt3=\frac1{2m}\langle \Xi^*_{bbc}|c^\dagger b|\Omega^*_{bbb}\rangle&=&\sqrt3\,\xi(1)
\end{eqnarray}
In the above equations, the left-most results follow from the
$c\leftrightarrow b$ $SU(2)$ symmetry, while the right-most ones are
obtained from Eqs.~(\ref{eq:tran2}), (\ref{eq:tran6}) and
(\ref{eq:tran8}). From Eqs.~(\ref{eq:norm1})--(\ref{eq:norm3}), we
deduce $\eta(1)=\chi(1)=\xi(1)=1$. For the actual quark masses one
expects deviations from this result as a consequence of a mismatch
between the initial and final wave functions.

\section{Decay width for a semileptonic $b\to c$ transition and HQSS constraints}
The total decay width for semileptonic $b\to c$ baryon transitions, is
given by
\bea 
\Gamma&=&|V_{cb}|^2
\frac{G_F^{\,2}}{8\pi^4}\frac{m'^2}{m} \int\sqrt{w^2-1}\, {\cal
L}^{\alpha\beta}(q) {\cal H}_{\alpha\beta}(v,k)\,dw 
\eea 
where $|V_{cb}|$ is the modulus of the corresponding
Cabibbo--Kobayashi--Maskawa (CKM) matrix element for a $b\to c$ quark
transition, for which we shall use $|V_{bc}|=0.0410$~\cite{pdg10}.
$G_F= 1.16637(1)\times 10^{-11}$\,MeV$^{-2}$~\cite{pdg10} is the Fermi
decay constant and $q=p-p'$. $w$ and $q^2$ are related by
$w=\frac{m^2+m'^2-q^2}{2mm'}$. In the decay, $w$ ranges from $w=1$,
corresponding to zero recoil of the final baryon, to a maximum value
given, neglecting the neutrino mass, by $w=w_{\rm max}= \frac{m^2 +
  m'^2-m_l^2}{2mm'}$, where $m_l$ is the final charged lepton mass.
Finally ${\cal L}^{\alpha\beta}(q)$ is the leptonic tensor after
integrating over the lepton momenta and ${\cal H}_{\alpha\beta}(v,k)$
is the hadronic tensor.

The leptonic tensor is given by
\bea
{\cal L}^{\alpha\beta}(q)=A(q^2)\,g^{\alpha\beta}+
B(q^2)\,\frac{q^\alpha q^\beta}{q^2}
\label{eq:lt}
\eea
where
\bea
A(q^2)=-\frac{I(q^2)}{6}\left(2q^2-m_l^2-\frac{m_l^4}{q^2}\right)\ ,\ \ 
B(q^2)=\frac{I(q^2)}{3}\left({q^2+m_l^2}-2\frac{m_l^4}{q^2}\right)
\end{eqnarray}
 with
\begin{eqnarray}
I(q^2)=\frac{\pi}{2q^2}(q^2-m_l^2)
\end{eqnarray}

The hadronic tensor reads
\begin{eqnarray}
{\cal H}^{\alpha\beta}(v,k) &=& \frac{1}{2J+1} \sum_{r,r'}  
 \big\langle B',v,k, r'\
\big| j^\alpha(0)\big| B, v,r\   \big\rangle 
\ \big\langle B',v,k, r'\ 
\big|j^\beta(0) \big|  B, v,r\  \big\rangle^*
\label{eq:wmunu}
\end{eqnarray}
with $J$ the initial baryon spin.
Baryonic states are normalized such that 
\bea \big\langle B, v', r'\
\, |\,B,v, r \  \big\rangle = 2E\,(2\pi)^3
\,\delta_{rr'}\, \delta (\vec{p}-\vec{p}^{\,\prime}) 
\eea 
with $E$
the baryon energy for three-momentum $\vec p$.
\subsection{HQSS constraints on semileptonic decay widths.}
\label{sect:hqsswidth}
For large quark masses and near zero recoil we can use the HQSS
results in Eqs.~(\ref{eq:tran1})--(\ref{eq:tran8}) 
to approximate the product ${\cal L}^{\alpha\beta}\,{\cal
H}_{\alpha\beta}$ by
\begin{itemize}
\item   $ccb\to ccc$ transitions
\begin{itemize}
\item $\Xi_{ccb}\to \Omega^*_{ccc}$
\begin{eqnarray}
\label{eq:lh1}
{\cal L}^{\alpha\beta}{\cal H}_{\alpha\beta}\approx
\frac{16}3\eta^2mm'(1+w)\bigg[-3A(q^2)+B(q^2)\bigg(\frac{(v'\cdot
q)^2}{q^2}-1\bigg)\bigg]
\end{eqnarray}
\item $\Xi^*_{ccb}\to \Omega^*_{ccc}$
\begin{eqnarray}
\label{eq:lh2}
{\cal L}^{\alpha\beta}{\cal H}_{\alpha\beta}\approx\frac{1}3\eta^2mm'\bigg[
-8A(q^2)w(1+2w^2)
+B(q^2)\bigg(-w(12+8w^2)+2\frac{(v\cdot q)(v'\cdot q)}{q^2}(20+8w^2)\bigg)\bigg]
\end{eqnarray}
\end{itemize}

\item  $bbc\to ccb$ transitions
\begin{itemize}
\item $\Xi_{bbc}\to \Xi_{ccb}$
\begin{eqnarray}
\label{eq:lh3}
{\cal L}^{\alpha\beta}{\cal H}_{\alpha\beta}\approx\frac49\chi^2mm'\bigg\{-A(q^2)(34w+32)+B(q^2)\bigg[
17\bigg(2\frac{(v\cdot q)(v'\cdot q)}{q^2}-w\bigg)-8\bigg]\bigg\}
\end{eqnarray}
\item $\Xi^*_{bbc}\to \Xi_{ccb}$
\begin{eqnarray}
\label{eq:lh4}
{\cal L}^{\alpha\beta}{\cal H}_{\alpha\beta}\approx
\frac89\chi^2mm'(1+w)\bigg[-3A(q^2)+B(q^2)\bigg(\frac{(v\cdot q)^2}{q^2}-1\bigg)\bigg]
\end{eqnarray}
\item $\Xi_{bbc}\to \Xi^*_{ccb}$
\begin{eqnarray}
\label{eq:lh5}
{\cal L}^{\alpha\beta}{\cal H}_{\alpha\beta}\approx
\frac{16}9\chi^2mm'(1+w)\bigg[-3A(q^2)+B(q^2)\bigg(\frac{(v'\cdot
q)^2}{q^2}-1\bigg)\bigg]
\end{eqnarray}
\item $\Xi^*_{bbc}\to \Xi^*_{ccb}$
\begin{eqnarray}
\label{eq:lh6}
{\cal L}^{\alpha\beta}{\cal H}_{\alpha\beta}\approx\frac{4}9\chi^2mm'\bigg[
-8A(q^2)w(1+2w^2)
+B(q^2)\bigg(-w(12+8w^2)+2\frac{(v\cdot q)(v'\cdot q)}{q^2}(20+8w^2)\bigg)\bigg]
\end{eqnarray}
\end{itemize}
\item  $bbb\to bbc$ transitions
\begin{itemize}
\item $\Omega^*_{bbb}\to \Xi_{bbc}$
\begin{eqnarray}
\label{eq:lh7}
{\cal L}^{\alpha\beta}{\cal H}_{\alpha\beta}\approx
\frac83\xi^2mm'(1+w)\bigg[-3A(q^2)+B(q^2)\bigg(\frac{(v\cdot q)^2}{q^2}-1\bigg)\bigg]
\end{eqnarray}
\item $\Omega^*_{bbb}\to \Xi^*_{bbc}$
\begin{eqnarray}
\label{eq:lh8}
{\cal L}^{\alpha\beta}{\cal H}_{\alpha\beta}\approx\frac13\xi^2mm'\bigg[
-8A(q^2)w(1+2w^2)
+B(q^2)\bigg(-w(12+8w^2)+2\frac{(v\cdot q)(v'\cdot q)}{q^2}(20+8w^2)\bigg)\bigg]
\end{eqnarray}
\end{itemize}
\end{itemize}

In the strict near zero recoil approximation, $\omega\approx 1$ or
equivalently $q^2$ very close to its maximum value $q^2_{\rm max}$, we
can approximate
\begin{equation}
\frac{(v\cdot q)^2}{q^2}\approx\frac{(v'\cdot q)(v\cdot
  q)}{q^2}\approx\frac{(v'\cdot q)^2}{q^2} \approx 1
  \label{eq:approx1}
\end{equation}
In addition, $A(q^2)\approx -B(q^2)$ near $q^2_{\rm max}$. To the
extent that the former approximations are good and further using
\begin{eqnarray}
m_{B_{bbc}}\approx m_{B^*_{bbc}}\ ;\ m_{B_{ccb}}\approx  m_{B^*_{ccb}}
\label{eq:approx_mass} 
\end{eqnarray}
we can make  approximate, but model independent, predictions for ratios of  
semileptonic $b\to c$ decay widths based in the above HQSS relations for 
${\cal L}^{\alpha\beta}{\cal H}_{\alpha\beta}$. We find
\begin{eqnarray}
\label{eq:r1}\frac{\Gamma(\Xi_{ccb}\to \Omega^*_{ccc})}
{\frac85\Gamma(\Xi^*_{ccb}\to \Omega^*_{ccc})}&\approx& 1
\\\nonumber\\
\label{eq:r2}\frac{\Gamma(\Xi_{bbc}\to \Xi_{ccb})}{\frac{25}8\Gamma(\Xi_{bbc}
\to \Xi^*_{ccb})}
&\approx&1\\
\label{eq:r3}\frac{2\Gamma(\Xi^*_{bbc}\to \Xi_{ccb})}{\Gamma(\Xi_{bbc}\to \Xi^*_{ccb})}&\approx&1\\
\label{eq:r4}\frac{\Gamma(\Xi^*_{bbc}\to
\Xi^*_{ccb})}{\frac52\Gamma(\Xi_{bbc}\to \Xi^*_{ccb})}
&\approx& 1
\\
\label{eq:r5}\frac{\Gamma(\Omega^*_{bbb}\to \Xi_{bbc})}{\frac45\Gamma(\Omega^*_{bbb}\to
\Xi^*_{bbc})}&\approx&1
\end{eqnarray}
These relations are similar to the ones we obtained in our former
study of doubly heavy baryons~\cite{Hernandez:2007qv} and from the
findings of this latter work, we expect them to hold at the level of
20\%. To estimate the decay widths themselves, we need to know the
Isgur-Wise functions $\eta(w),\chi(w)$ and $\xi(w)$. In the next
section we will use a nonrelativistic constituent quark model for this
purpose.

\section{Nonrelativistic quark model evaluation of the Isgur-Wise functions and
decay widths}
In this section we shall obtain, within the nonrelativistic quark
model and using the AL1 interquark potential of
Refs.~\cite{semay94,silvestre96}, the wave functions of the heavy
baryons involved in this study. With those wave functions we can
evaluate the Isgur-Wise functions and estimate the baryon semileptonic
$b\to c$ decay widths.

The wave functions have the general form
\bea
\Psi_{\alpha_1\alpha_2\alpha_3}=\delta_{f_1h}\delta_{f_2h}
\delta_{f_3h'}\frac{\epsilon_{c_1c_2c_3}}{\sqrt{3!}}\Phi(r_1,r_2,r_{12})
(1/2,1/2,1;s_1,s_2,s_1+s_2)(1,1/2,J;s_1+s_2,s_3,M) 
\eea 
where $\alpha_j$ represents the spin (s), flavour (f) and colour (c)
quantum numbers of the $j$-th quark. The two first quarks have the
same flavour $h$, while the third quark has flavour $h'$, which could
be also the same as the one of the first two.
$\epsilon_{c_1c_2c_3}/\sqrt{3!}$ is the fully antisymmetric colour
wave function and the $(j_1,j_2,j;m_1,m_2,m)$ are $SU(2)$
Clebsch-Gordan coefficients. $J$ is the total spin of the baryon. As
we are interested only in spin $1/2$ or $3/2$ ground state baryons,
the total orbital angular momentum is $L=0$. Thus, the orbital part of
the wave function can only depend on the modulus of the relative
distances between the quarks. Here we use $r_1, r_2$ which are the
relative distances between quark three and quarks one and two
respectively, and $r_{12}$ which is the relative distance between the
first two quarks. Following our works on single and double heavy
baryons~\cite{Albertus:2003sx,Albertus:2006wb} we shall use a
variational ansatz to solve the three-body problem. We write the
orbital wave functions as the product of three functions, each one
depending on just one of the three variables $r_1,r_2,r_{12}$, i.e.
\bea
\Phi(r_1,r_2,r_{12})=\phi_{hh'}(r_1)\phi_{hh'}(r_2)\phi_{hh}(r_{12})
\eea 
For each of the $\phi$ functions above we take an expression of the
form\footnote{We use four
gaussians in the present approach. We have checked that by
increasing the number of gaussians, the variational baryon
masses change in less than 5 MeV.}  
\bea \phi(r)=\sum_{j=1}^4a_je^{-b_j^2(r+d_j)^2}\hspace{1cm}
(a_1=1).  \eea 
The variational parameters are fixed by minimizing the
energy and the overall normalization is fixed at the end of the
calculation. The results we get for the masses are given in
Table~\ref{tab:masses}, where we also compare them to the ones
obtained in Ref.~\cite{silvestre96} using the same potential but
solving Faddeev equations. The agreement between the two approaches is
very good. 

There is a recent estimate~\cite{Meinel:2010pw} of the mass of the
triply-heavy baryon $bbb$ obtained in lattice QCD with $2{+}1$ flavours
of light sea quarks. Our result compares with it rather well. Our
predictions are also in a reasonable agreement with those obtained
within the BM, RTQM and LO pNRQCD evaluations of
Refs.~\cite{Hasenfratz:1980ka}, \cite{Martynenko:2007je} and
\cite{Jia:2006gw}, respectively. The QCDSR masses calculated in
\cite{Zhang:2009re} come out systematically much smaller than ours,
while those obtained in the NRCQM of
Ref.~\cite{Roberts:2007ni} are significantly larger than our
predictions.

Before computing the Isgur-Wise functions that govern the semileptonic
decays of the triply baryons, we would like to devote a few words to
discussing the confinement potential in these systems. In
phenomenological constituent quark models, such as the AL1 potential
used here, the confinement potential for baryons is usually obtained
from the two body forces that describe the dynamics of each quark
pair. However, the lattice QCD simulations carried out in
Refs.~\cite{Takahashi:2000te,Takahashi:2002bw} seem to indicate that
in a three static quark system, confinement is a genuine three-body
effect. Changes in the masses due to the use of one or other of these
approaches are studied in the next subsection.

\begin{table}
\begin{tabular}{lcc|cccccccccc}\hline\hline
&This work&\cite{silvestre96}&\cite{Meinel:2010pw}&
  \cite{Hasenfratz:1980ka}&\cite{Bjorken}  & \cite{Vijande:2004at} 
& \cite{Jia:2006gw}& \cite{Roberts:2007ni} &\cite{Martynenko:2007je}
    & \cite{Guo:2008he} &
  \cite{Zhang:2009re} \\
&Variational&Faddeev&LQCD & BM& &NRCQM & 
 LO pNRQCD & NRCQM& RTQM & Regge & QCDSR\\\hline
$m_{\Omega^*_{bbb}}$&14398&14398& $14371 \pm 12$ & 14300 & $14760 \pm
  180$ &--  & $14370 \pm 80$&14834&    14569 &-- & $13280 \pm 100$\\
$m_{\Xi^*_{bbc}}$&11245&--&-- &11200 & $11480 \pm 120$ &--   &$11190
  \pm 80$& 11554& 11287 & -- &$10540 \pm 110$\\
$m_{\Xi_{bbc}}$&11214&11217&--&-- &-- & -- &$11190
  \pm 80$ &11535&  11280&-- &$10300 \pm 100$\\
$m_{\Xi^*_{ccb}}$&8046&--&--&8030 & $8200 \pm 90$& --  &$7980
  \pm 70$& 8265&  8025&-- &$7450 \pm 160$\\
$m_{\Xi_{ccb}}$&8018&8019&--&-- &-- &--  &$7980 \pm 70$& 8245&  8018&--& $7410 \pm 130$\\
$m_{\Omega^*_{ccc}}$&4799&4799&--&4790 &$4925\pm 90$ & 4632 & $4760 \pm
  60$&4965&  4803 & $4819 \pm 7$ & $4670 \pm 150$\\
\hline\hline
\end{tabular}
\caption{Masses (in MeV) of the triply heavy baryons obtained with the
AL1 potential of Refs.~\cite{semay94,silvestre96} using our
variational approach. For comparison we also show the results from the
Faddeev calculation performed in Ref.~\cite{silvestre96} using the
same potential.  Predicted masses within other theoretical approaches
are also compiled. Hyperfine splitting is neglected in
\cite{Jia:2006gw}. }
\label{tab:masses}
\end{table}

\subsection{$\Delta$-shaped versus $Y$-shaped potential}
\begin{figure}
\resizebox{8.cm}{!}{\includegraphics{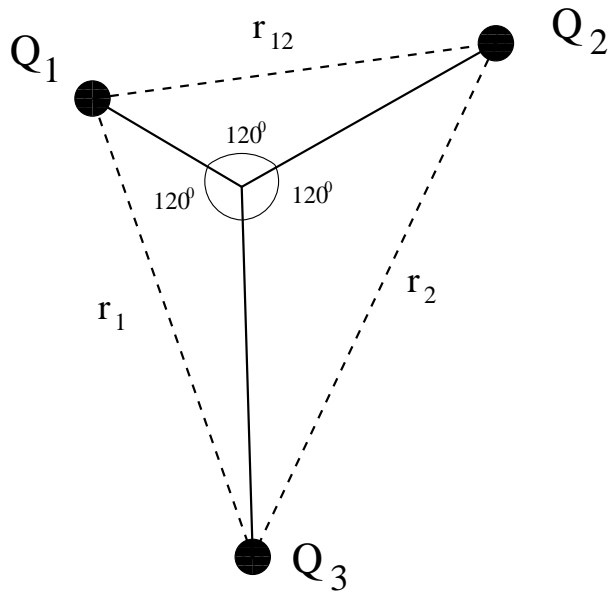}}
\caption{Illustration of the $\Delta$- and $Y$-shaped confinement potentials.}
  \label{fig:y2}
\end{figure}

LQCD results for the static quark-antiquark ground-state
potential~\cite{Bali:1994de} are well described by a dependence
\be
-\frac{A}{r}+\sigma r+C
\ee
which contains the sum of the short distance Coulomb one-gluon
exchange (OGE) term plus the confining long distance flux-tube
contribution. Most phenomenological models assume such a dependence
and fit the $A,\sigma$ and $C$ parameters to the meson spectrum. This
is for instance the case of the AL1 potential that we use. When going
to the quark-quark sector, a factor of 1/2, assumed to come from an
overall colour $\vec\lambda\cdot\vec\lambda$ dependence\footnote{QCD
  predicts exactly this colour factor for the OGE term.}, is added to
the interaction. The resulting potential in baryons is thus obtained
as the sum of two-body terms. For the confining part one is summing
over the three sides of a triangle with the quarks located at its
vertices [$\sigma (r_1+r_2+r_{12})/2$], leading to the name of
$\Delta$-shaped potential (see Figure~\ref{fig:y2}). This picture
works very well from a phenomenological point of view and one gets a
good description of the light and single heavy baryon spectrum once
the parameters have been fixed in the corresponding meson sector.

As mentioned above, the $3$ quark static potential has been directly
measured on the lattice in
Refs.~\cite{Takahashi:2000te,Takahashi:2002bw}. A good fit to the
lattice data was obtained assuming a picture, similar to the one
described above, in which the potential has a short-distance Coulomb
OGE part plus a long-distance flux-tube part
\be
-A_{3q}\big(\frac{1}{r_1}+\frac{1}{r_2}+\frac{1}{r_{12}}\big)
+\sigma_{3q}L_{\rm min}+C_{3q} \label{eq:3body}
\ee 
where $L_{\rm min}$ is the minimal value of the total length of the
colour flux tubes linking the three quarks. The flux tubes adopt a $Y$
shape (See Fig.~\ref{fig:y2}), hence the name $Y$-shaped potential.
This is in agreement with the picture that emerges from the QCD BM
 calculations carried out in Ref.~\cite{Hasenfratz:1980ka}.
Indeed the $\Delta$ and $Y$ nomenclature was already used in this
pioneering work of 1980. In terms of the $r_1,r_2$ and $r_{12}$
interquark distances one has
\be L_{\rm
min}^2=\frac12(r_1^2+r_2^2+r_{12}^2)+\frac{\sqrt3}{2}
\sqrt{-\lambda(r^2_1,r^2_2,r^2_{12})} 
\ee 
when none of the angles of the three quark triangle exceeds $2\pi/3$
and where $\lambda(a,b,c)=a^2+b^2+c^2-2ab-2ac-2bc$. If one of the
triangle's angles exceeds $2\pi/3$ then $L_{\rm min}$ is just given by
\be
L_{\rm min}=r_1+r_2+r_{12}-{\rm max}(r_1,r_2,r_{12})
\ee
Comparing this fit with the one for the quark-antiquark potential they
found that $\sigma_{3q}\approx\sigma$, $A_{3q}\approx\frac12A$ and
$C_{3q}\approx \frac32 C$. Thus, leaving out the confinement piece,
one could approximate the $3$ quark potential by the sum of three
two-body quark-quark terms. For the confining part, the sum of three
two-body quark-quark terms is always smaller than the three body force
obtained from lattice QCD data. Actually, one has $ (r_1+r_2+r_{12})/2
\le L_{\rm min} \le (r_1+r_2+r_{12})/\sqrt 3$, which might induce
changes of around 15\% at most in this part of the potential. Indeed,
in Ref.~\cite{Takahashi:2000te} the lattice data were also fitted to
the sum of the three quark-quark potentials\footnote{The fit is worse
  than that obtained when the functional form of Eq.~(\ref{eq:3body})
  is used.} and it was found that a slightly larger value for the
confinement coefficient (0.53 $\sigma$ vs $\sigma$/2) was required.
The ad-hoc factor of $1/2$ introduced in quark potentials when going
from the mesons to baryons is understood here as a geometrical effect
rather than as a colour factor as it is usually presented.

To be more quantitative, we have computed the triply heavy baryon
masses also with a $Y$-shaped confinement potential. To that end, we
have taken the AL1 potential used before, and have replaced the
$\sigma(r_1+r_2+r_{12})/2$ term by $\sigma L_{\rm min}$. Results are
presented in Table~\ref{tab:DeltaY}. There we also compare with the
masses obtained previously using the AL1 potential. We see a small
increase in the masses of roughly $26$, $34$, $40$ and
$48\,\mathrm{MeV}$ for the $bbb,\,bbc,\,ccb$ and $ccc$ systems
respectively. Effects here are similar to those due to the hyperfine
splitting. Future and precise measurements of the masses might help to
shed light on the exact nature of the confinement potential in the
baryon sector.

The corrections to the ratios of decay widths, that would be computed
in the subsections below, are even smaller, as expected from
perturbation theory, since changes in the wave functions arise at
second order.
\begin{table}
\begin{tabular}{lcc}\hline\hline
&\multicolumn{2}{c}{This work}\\
&$\Delta$-shaped potential&$Y$-shaped potential\\\hline
$m_{\Omega^*_{bbb}}$&14398&14424\\
$m_{\Xi^*_{bbc}}$&11245&11281\\
$m_{\Xi_{bbc}}$&11214&11247\\
$m_{\Xi^*_{ccb}}$&8046&8087\\
$m_{\Xi_{ccb}}$&8018&8058\\
$m_{\Omega^*_{ccc}}$&4799&4847\vspace{.1cm}\\
\hline\hline
\end{tabular}
\caption{First column: Masses (in MeV) of the triply heavy baryons
obtained with the AL1 potential of Refs.~\cite{semay94,silvestre96}
using our variational approach. Second column: The same by
substituting $\sigma(r_1+r_2+r_{12})/2$ by $\sigma L_{\rm min}$
in the AL1 potential.}
\label{tab:DeltaY}
\end{table}
\subsection{Isgur-Wise functions}
\begin{figure}
\resizebox{12.cm}{!}{\includegraphics{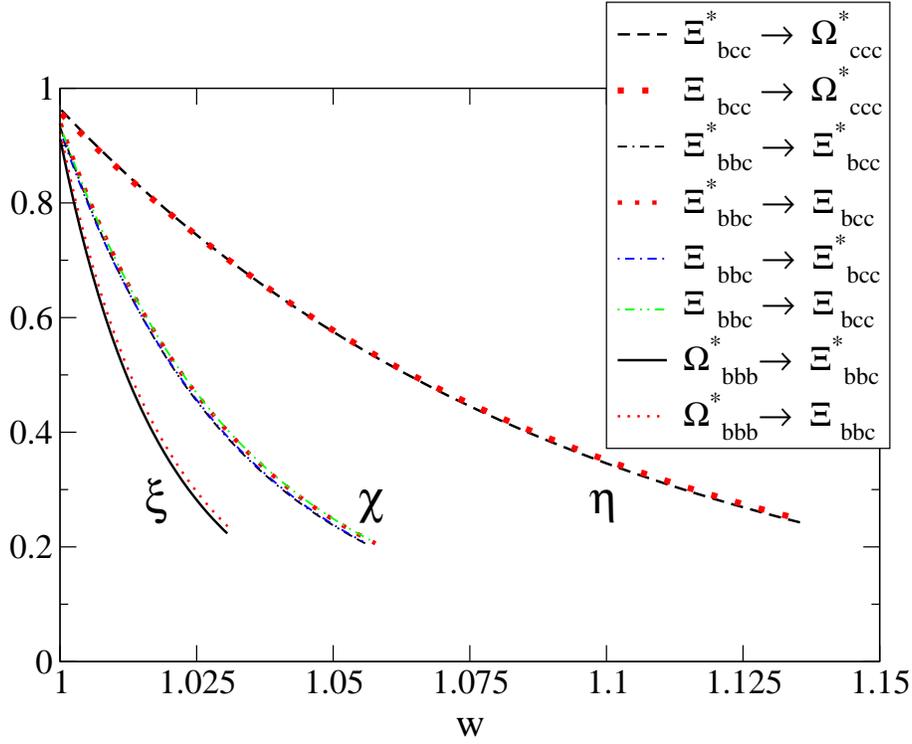}}
\vspace{0.4cm}
\caption{Overlap functions for $b\to c$ semileptonic decays of
  triply-heavy baryons obtained in a nonrelativistic quark model. The
  functions fall into three families, consistent with heavy quark spin
  symmetry.}
  \label{fig:iw}
\end{figure}

To evaluate the Isgur-Wise functions we follow our work in
Ref.~\cite{Albertus:2004wj} and write\footnote{Note when the initial
baryon is at rest, $w=\frac{E'}{m'}$ is just a function of $|\vec
q\,|$.}
\bea
{\eta_{\Xi^{(*)}}}(|\vec q\,|)&=&  
\int d^3r_1 d^3r_2 \, e^{{\rm i}\vec{q}\cdot\left(m_c\vec{r}_1+
m_{c}\vec{r}_2 \right)/\overline{M}_{ccc}}
[\Psi_{\Omega^*_{ccc}}(r_1,r_2,r_{12})]^*\ 
\Psi_{\Xi^{(*)}_{ccb}}(r_1,r_2,r_{12})\\
{\chi_{\Xi^{(*)}\to \Xi, \Xi^{*}}}(|\vec q\,|)&=&  
\int d^3r_1 d^3r_2 \, e^{-{\rm i}\vec{q}\cdot\left(m_b\vec r_{12}+m_c\vec{r}_1
\right)/\overline{M}_{ccb}}
[\Psi_{\Xi^{(*)}_{ccb}}(r_{12},r_2,r_1)]^*\ 
\Psi_{\Xi^{(*)}_{bbc}}(r_1,r_2,r_{12})\\ 
{\xi_{\Xi^{(*)}}}(|\vec q\,|)&=&  
\int d^3r_1 d^3r_2 \, e^{{\rm i}\vec{q}\cdot\left(m_b\vec{r}_1+
m_{b}\vec{r}_2 \right)/\overline{M}_{bbc}}
[\Psi_{\Xi^{(*)}_{bbc}}(r_1,r_2,r_{12})]^*\ 
\Psi_{\Omega^*_{bbb}}(r_1,r_2,r_{12})
\eea
where $\overline{M}_{hhh'}=m_h+m_h+m_{h'}$. In Fig.~\ref{fig:iw}, we
display the eight overlap functions obtained from each of the decays
examined here. We see that as predicted by HQSS in
Eqs.~(\ref{eq:tran1})--(\ref{eq:tran8}) they reduce to only three
independent ones in very good approximation. In the equal mass case
they would be equal to one at zero recoil ($|\vec q\,|=0$). For finite
masses we see they deviate slightly from $1$ at zero recoil owing to
the mismatch between the initial and final wave functions. Note that
$w= 1+ v\cdot k /m'$, and thus as $w$ departs from 1, $v\cdot k /m'$
increases. To obtain the relations of
Eq.~(\ref{eq:tran1})--(\ref{eq:tran8}) all $v\cdot k /m'$ corrections
were neglected. Thus, the Isgur-Wise (overlap) functions depicted in
Fig~\ref{fig:iw} would provide a poorer description of the weak
transition matrix elements as $w$ deviates from the zero recoil point.
The largest corrections are expected for the $bcc \to ccc$
transitions, related to the $\eta-$type Isgur-Wise functions in
Fig~\ref{fig:iw}, for which $w_{\rm max}\approx 1.125$. For this case
at the $q^2=0$ end of the phase space, $v\cdot k /m'$ becomes of order
$1/8$. Thus, in this region, approximating the full amplitude (weak
matrix element) by means of the $\eta(w)$ Isgur-Wise function could be
subject to uncertainties of order of $15$--$25\%$. For $bbc\to bcc$ or
$bbb\to bbc$ transitions, the $\chi-$ and $\xi-$ Isgur-Wise functions
should provide more accurate estimates of the transition weak matrix
elements for the whole available phase space, since in those cases
$w_{\rm max}$ is only about $1.06$ and $1.03$, respectively.

In the next subsection, we will make use of the Isgur-Wise functions
of Fig~\ref{fig:iw} to estimate the decay widths. This should be quite
accurate, even for the $bcc\to ccc$ transitions, since, as we will
see, the differential decay width distribution peaks very close to the
zero recoil point and hence far from the end point of the spectrum
$w=w_{\rm max}$. Indeed, for $bcc\to ccc$ transitions, the
distribution takes its maximum value well below $w=1.05$ (see left
upper panel of Fig.~\ref{fig:dgdw}).
\subsection{Decay widths}
\begin{figure}
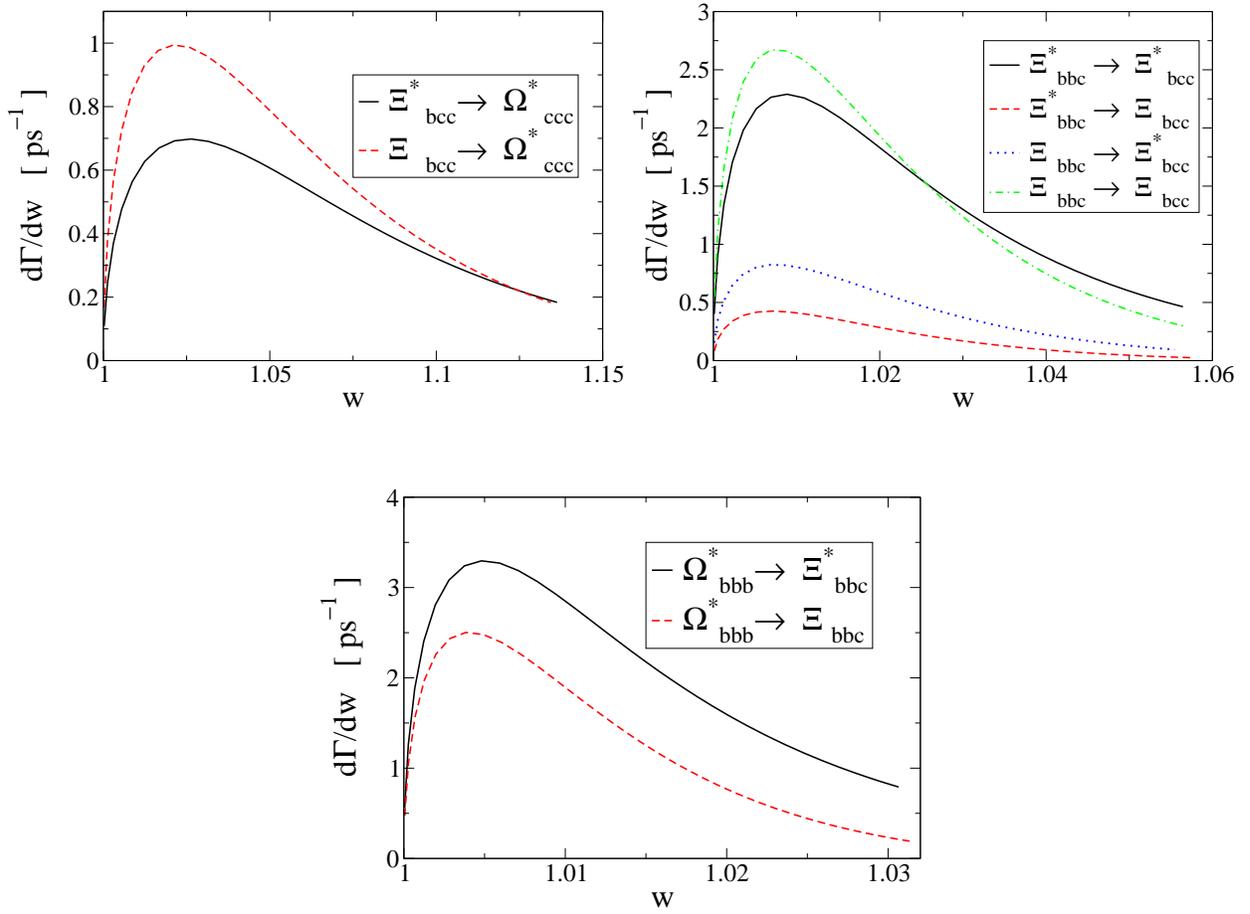

\resizebox{8.2cm}{!}{\includegraphics{dgdw_bcc.eps}}
\resizebox{8.cm}{!}{\includegraphics{dgdw_bbc.eps}}\vspace{1cm}\\
\resizebox{8.cm}{!}{\includegraphics{dgdw_bbb.eps}}
\caption{Estimated $\frac{d\Gamma}{dw}$ differential 
decay widths in $ps^{-1}$ for the different transitions considered.}
  \label{fig:dgdw}
\end{figure}

We now use the HQSS approximate expressions in
Eqs.~(\ref{eq:lh1})--(\ref{eq:lh8}) to estimate the decay widths. The
results for the differential distributions are shown in
Fig.~\ref{fig:dgdw} and the integrated decay widths are compiled in
Table~\ref{tab:dw}. We see the ratios in Eqs.~(\ref{eq:r2}) and
(\ref{eq:r3}) are satisfied at the level of 5.5\% and 3.4\%
respectively whereas the ratios in Eqs.~(\ref{eq:r1}), (\ref{eq:r4})
and (\ref{eq:r5}) are good only at the level of 20-30\%. It is clear
the relations in Eqs.~(\ref{eq:r1})--(\ref{eq:r5}) can only be
approximate. First, the strict zero recoil point is forbidden by phase
space, and second $q^2$ changes rapidly from its maximum value of
$(m-m')^2$ at $w=1$ to its minimum value of $m_l^2$ at $w_{max}$ which
makes the approximation in Eq.(\ref{eq:approx1}) not good
enough\footnote{Note, as pointed out in Ref.~\cite{Hernandez:2007qv},
  the quantities $(v\cdot q)^2/q^2$, $(v'\cdot q) (v\cdot q)/q^2$ and
  $(v'\cdot q)^2/q^2$ which are all equal to 1 near zero recoil,
  quickly deviate from 1 because of the $q^2$ factor in the
  denominator.}. 

What one sees when looking at the differential decay widths in
Fig.~\ref{fig:dgdw} is that these distributions peak in each case in
the lower part of the allowed $w$ region, about $1.005$, $1.009$ and
$1.025$ for $bbb,bbc$ and $ccb$ decays respectively, quite close to
the zero recoil point. In these circumstances
one can relax the strict approximation in Eq.(\ref{eq:approx1}) and
use instead~\cite{Hernandez:2007qv}
\begin{eqnarray}
\frac{(v\cdot q)^2}{q^2}\approx\frac{(v\cdot q)(v'\cdot q)}{q^2}\approx
\frac{(v'\cdot q)^2}{q^2}
\end{eqnarray}
which should be reasonable near the maximum of the differential decay
width, since we can still use $w\approx 1$. We can also use
$B(q^2)\approx-A(q^2)$ and the approximate equality of masses in
Eq.(\ref{eq:approx_mass}). One can now only make the following two
model independent predictions
\bea
 \frac{2\Gamma(\Xi^*_{bbc}\to
  \Xi_{ccb})}{\Gamma(\Xi_{bbc}\to \Xi^*_{ccb})}&\approx&
1\\ \frac{\Gamma(\Xi^*_{bbc}\to\Xi^*_{ccb})}{4\Gamma(\Xi_{bbc}\to\Xi_{ccb})-
  10\Gamma(\Xi_{bbc}\to\Xi^*_{ccb})}&\approx& 1
\eea
which we see are good at the level of 3.4\% and 0.25\% respectively.
\begin{table}
\begin{tabular}{cc}\hline\hline
$B\to B'e \bar\nu_e$&\hspace{.5cm}$\Gamma\ [\,{\rm ps}^{-1}]$\hspace*{.5cm}\\\hline
$\Xi_{ccb}\to\Omega^*_{ccc}\,e\bar\nu_e$&$8.01\times 10^{-2}$\\
$\Xi^*_{ccb}\to\Omega^*_{ccc}\,e\bar\nu_e$&$6.28\times 10^{-2}$\\
$\Xi_{bbc}\to\Xi_{ccb}\,e\bar\nu_e$&$7.98\times 10^{-2}$\\
$\Xi_{bbc}\to\Xi^*_{ccb}\,e\bar\nu_e$&$2.42\times 10^{-2}$\\
$\Xi^*_{bbc}\to\Xi_{ccb}\,e\bar\nu_e$&$1.17\times 10^{-2}$\\
$\Xi^*_{bbc}\to\Xi^*_{ccb}\,e\bar\nu_e$&$7.74\times 10^{-2}$\\
$\Omega^*_{bbb}\to\Xi_{bbc}\,e\bar\nu_e$&$3.95\times 10^{-2}$\\
$\Omega^*_{bbb}\to\Xi^*_{bbc}\,e\bar\nu_e$&$6.34\times 10^{-2}$\\
\hline\hline
\end{tabular}
\caption{Estimated decay widths  in units of ${\rm ps}^{-1}$. We use
$|V_{bc}|=0.0410$. Similar results are obtained for
$\mu\bar\nu_\mu$ leptons in the final state.}
\label{tab:dw}
\end{table}

\section{Conclusions}

We have studied the $b\to c$ semileptonic decays of the lowest lying
triply heavy ($Q_1Q_2Q_3$, with $Q_i=b,c$) baryons in the limit
$m_b,m_c \gg \Lambda_{\rm QCD}$ and close to the zero-recoil point.
The separate heavy quark spin symmetries strongly constrain the matrix
elements, leading to single form factors for all these decays. We have
obtained these HQSS relations for the first time. Lattice QCD
simulations work best near the zero-recoil point and thus are well
suited to check the validity of our results.

We have used a NRCQM, adjusted to
the meson spectrum, to predict the masses of these triply heavy
baryons by using a simple variational scheme. Results for masses
compare rather well with some previous Faddeev and LQCD estimates. We
have also obtained masses by using a lattice QCD inspired three-body
confinement potential. The variational wave functions have been
employed to compute the overlap integrals needed to evaluate the
relevant Isgur-Wise functions that describe these decays. We have
checked that our calculations are consistent with HQSS and have used
them to estimate the semileptonic decay widths.

We have in addition made approximate, but model independent,
predictions for ratios of semileptonic $b\to c$ decay widths based on
the HQSS relations derived here, which we expect to be accurately
fulfilled.

\begin{acknowledgments}
  This research was supported by DGI and FEDER funds, under contracts
  FIS2011-28853-C02-02, FPA2010-21750-C02-02, and the Spanish
  Consolider-Ingenio 2010 Programme CPAN (CSD2007-00042),  by Generalitat
  Valenciana under contract PROMETEO/20090090 and by the EU
  HadronPhysics2 project, grant agreement n.~227431. 
\end{acknowledgments}


\end{document}